\documentclass{aa}
\usepackage{epsfig,graphicx,natbib,url,twoopt}
\usepackage[varg]{txfonts}
\usepackage{hyperref}          
\usepackage[usenames]{color}
\usepackage{pdfcomment,acronym}      
\hypersetup{
  colorlinks=true,  
  urlcolor=blue,    
  linkcolor=red     
}
\newcount\markrevised

\def\RR #1\par{\noindent{\small \color{red} $\sharp$RR #1}\\[1ex]}
 
\bibpunct{(}{)}{;}{a}{}{,}    
\makeatletter
 \newcommandtwoopt{\citeads}[3][][]{%
   \nonstopmode
   \href{http://adsabs.harvard.edu/abs/#3}%
        {\def\hyper@linkstart##1##2{}%
         \let\hyper@linkend\@empty\citealp[#1][#2]{#3}}
   \biblink{#3}{\href{http://adsabs.harvard.edu/abs/#3}{ADS}}%
   \errorstopmode}            
 \newcommandtwoopt{\citepads}[3][][]{%
   \nonstopmode
   \href{http://adsabs.harvard.edu/abs/#3}%
        {\def\hyper@linkstart##1##2{}%
         \let\hyper@linkend\@empty\citep[#1][#2]{#3}}
   \biblink{#3}{\href{http://adsabs.harvard.edu/abs/#3}{ADS}}%
   \errorstopmode}            
 \newcommandtwoopt{\citetads}[3][][]{%
   \nonstopmode
   \href{http://adsabs.harvard.edu/abs/#3}%
        {\def\hyper@linkstart##1##2{}%
         \let\hyper@linkend\@empty\citet[#1][#2]{#3}}
   \biblink{#3}{\href{http://adsabs.harvard.edu/abs/#3}{ADS}}%
   \errorstopmode}            
 \newcommandtwoopt{\citeyearads}[3][][]{%
   \nonstopmode
   \href{http://adsabs.harvard.edu/abs/#3}%
        {\def\hyper@linkstart##1##2{}%
         \let\hyper@linkend\@empty\citeyear[#1][#2]{#3}}
   \biblink{#3}{\href{http://adsabs.harvard.edu/abs/#3}{ADS}}%
   \errorstopmode}            
\makeatother

\def\linkadspage#1#2#3{\href{http://adsabs.harvard.edu/cgi-bin/nph-data_query?bibcode=#1\&link_type=ARTICLE\&d_key=AST\#page=#2}{#3}}

\newcount\longrefs
\def\aap{\ifnum\longrefs=1 {Astron.\ Astrophys.}\else 
                           {A\hbox{\rm \&}A}\fi}
\def\aapr{\ifnum\longrefs=1 {Astron.\ Astrophys.\ Rev.}\else 
                            {A\hbox{\rm \&}AR}\fi}
\def\aaps{\ifnum\longrefs=1 {Astron.\ Astrophys.\ Suppl.}\else 
                            {A\hbox{\rm \&}A Suppl.}\fi}
\def\actaa{\ifnum\longrefs=1 {Acta Astronomica}\else
                            {Acta Astron.}\fi}
\def\aipcs{\ifnum\longrefs=1 {Am.\ Inst.\ Phys.\ Conf.\ Series}\else
                             {AIP Conf.\ Ser.}\fi}
\def\aj{\ifnum\longrefs=1 {Astron.\ J.}\else 
                          {AJ}\fi} 
\def\ao{\ifnum\longrefs=1 {Applied Optics}\else 
                           {Appl.\ Opt.}\fi} 
\def\aspcs{\ifnum\longrefs=1 {Astron.\ Soc.\ Pacific Conf.\ Series}\else 
                           {ASP Conf.\ Ser.}\fi} 
\def\apj{\ifnum\longrefs=1 {Astrophys.\ J.}\else 
                           {ApJ}\fi} 
\def\apjl{\ifnum\longrefs=1 {Astrophys.\ J. Lett.}\else 
                            {ApJL}\fi} 
\def\aplett{\ifnum\longrefs=1 {Astrophys.\ J. Lett.}\else 
                            {ApJ}\fi} 
\def\apjs{\ifnum\longrefs=1 {Astrophys.\ J. Suppl.}\else 
                            {ApJS}\fi}
\def\apss{\ifnum\longrefs=1 {Astrophys.\ and Space Science}\else 
                            {Astrophys.\ Space Sci.}\fi}
\def\araa{\ifnum\longrefs=1 {Ann.\ Rev.\ Astron.\ Astrophys.}\else 
                            {ARA\hbox{\rm \&}A}\fi}
\def\azh{\ifnum\longrefs=1 {Astronomicheskii Zhurnal}\else 
                            {Astron.\ Zhur.}\fi}
\def\baas{\ifnum\longrefs=1 {Bull.\ Am.\ Astron.\ Soc.}\else 
                            {BAAS}\fi}
\def\bain{\ifnum\longrefs=1 {Bull.\ Astronom.\ Institutes Netherlands}\else
                            {Bull.\ Astr.\ Inst.\ Neth.}\fi}
\def\cjaa{\ifnum\longrefs=1 {Chinese Jour.\ Astron.\ Astrophys.}\else 
                            {Chin.\ J.\ A\&A}\fi}
\def\gca{\ifnum\longrefs=1 {Geochim.\ Cosmochim.\ Acta}\else 
                           {Geochim.\ Cosmochim.\ Acta}\fi}
\def\grl{\ifnum\longrefs=1 {Geophys.\ Res.\ Lett.}\else 
                           {Geoph.\ Res.\ Lett.}\fi}
\def\iaucirc{\ifnum\longrefs=1 {IAU Circulars}\else 
                          {IAU Circ.}\fi}
\def\icarus{\ifnum\longrefs=1 {Icarus}\else 
                          {Icarus}\fi}
\def\ip{\ifnum\longrefs=1 {in press}\else 
                          {in press}\fi}
\def\jcap{\ifnum\longrefs=1 {Jour.\ Cosmology Astropart.\ Phys.}\else 
                          {JCAP}\fi}
\def\jgr{\ifnum\longrefs=1 {J.\ Geophys.\ Res.}\else 
                           {J.\ Geophys.\ Res.}\fi}  
\def\jrasc{\ifnum\longrefs=1 {J.\ Royal Astron.\ Soc.\ Canada}\else 
                           {JRAS Can.}\fi}  
\def\memsai{\ifnum\longrefs=1 {Mem.~Soc.~Astron.~Italiana}\else
                              {MmSAI}\fi}
\def\mnras{\ifnum\longrefs=1 {Mon.\ Not.\ Roy.\ Astron.\ Soc.}\else 
                             {MNRAS}\fi} 
\def\na{\ifnum\longrefs=1 {New Astronomy}\else 
                           {New Astron.}\fi}
\def\nar{\ifnum\longrefs=1 {New Astronomy rev.}\else 
                           {New Astron.\ Rev.}\fi}
\def\nat{\ifnum\longrefs=1 {Nature}\else 
                           {Nat}\fi}
\def\pasa{\ifnum\longrefs=1 {Pub.\ Astron.\ Soc.\ Australia}\else 
                            {PASA}\fi} 
\def\pasj{\ifnum\longrefs=1 {Pub.\ Astron.\ Soc.\ Japan}\else 
                            {PASJ}\fi} 
\def\pasp{\ifnum\longrefs=1 {Pub.\ Astron.\ Soc.\ Pacific}\else 
                            {PASP}\fi} 
\def\physscr{\ifnum\longrefs=1 {Physica Scripta}\else 
                            {Phys.\ Scrip.}\fi} 
\def\planss{\ifnum\longrefs=1 {Planetary \& Space Science}\else 
                            {Plan. \& Space Sci.}\fi} 
\def\procspie{\ifnum\longrefs=1 {Proc.\ SPIE}\else 
                            {Proc.\ SPIE}\fi} 
\def\qjras{\ifnum\longrefs=1 {Quarterly J.\ Royal Astron.\ Soc.}\else 
                            {QJRAS}\fi} 
\def\rmxaa{\ifnum\longrefs=1 {Revista Mexicana de Astron.\ y Astrofys.}\else 
                            {RMxAA}\fi} 
\def\sa{\ifnum\longrefs=1 {Soviet Astron..}\else 
                               {Sov.\ Astron.}\fi}
\def\skytel{\ifnum\longrefs=1 {Sky \& Telescope}\else 
                            {Sky \& Tel.}\fi} 
\def\solphys{\ifnum\longrefs=1 {Solar Phys.}\else 
                               {SoPh}\fi}
\def\sovast{\ifnum\longrefs=1 {Soviet Astronomy}\else 
                               {Sov.\ Ast.}\fi}
\def\ssr{\ifnum\longrefs=1 {Space Science Rev.}\else 
                               {Space\ Sci.\ Rev.}\fi}
\def\zap{\ifnum\longrefs=1 {Zeitschr.\ f.\ Astrophysik}\else
                               {Z.\ Astrophys.}\fi}

\makeatletter
\newcommand{\bibnote}[2]{\global\@namedef{#1note}{#2}}
\newcommand{\biblink}[2]{\global\@namedef{#1link}{#2}}
\makeatother

\def\wlchianti#1{\href{http://www.chiantidatabase.org}{#1}}
\def\wldotmovies#1{\href{http://www.staff.science.uu.nl/~rutte101/dot/albums/movies/album.html}{#1}}
\def\wlRRcourses#1{\href{http://www.staff.science.uu.nl/~rutte101/Astronomy_course.html}{#1}}
\def\wlRRidl#1{\href{http://www.staff.science.uu.nl/~rutte101/Recipes_IDL.html}{#1}}
\def\wlRRssx#1{\href{http://www.staff.science.uu.nl/~rutte101/rrweb/rjr-edu/lectures/rutten_ssx_lec.pdf}{#1}}

\def\wlsolarsoft#1{\href{http://www.lmsal.com/solarsoft}{#1}}

\newacro{AA}{Astronomy \& Astrophysics}  
\newacro{ADS}{Astrophysics Data System}
\newacro{AIA}{Atmospheric Imaging Assembly}
\newacro{ALMA}{Atacama Large Millimeter/submillimeter Array}
\newacro{AO}{adaptive optics}
\newacro{ApJ}{Astrophysical Journal}
\newacro{AR}{active region}
\newacro{BFI}{Broad-band Filter Imager}
\newacro{CE}{coronal equilibrium}
\newacro{CfA}{Center for Astrophysics}
\newacro{CME}{coronal mass ejection}
\newacro{CRD}{complete redistribution}
\newacro{CRISP}{CRisp Imaging SpectroPolarimeter}
\newacro{CRISPEX}{CRisp SPectral EXplorer}
\newacro{CS}{coherent scattering}
\newacro{DEM}{Differential Emission Measure}
\newacro{DF}{dynamic fibril}
\newacro{DKIST}{Daniel K. Inouye Solar Telescope}
\newacro{DLR}{Deutsches Zentrum f\"ur Luft- und Raumfahrt}
\newacro{DOT}{Dutch Open Telescope}
\newacro{DST}{Richard B. Dunn Solar Telescope}   
\newacro{EB}{Ellerman bomb}
\newacro{EDP}{\'{E}dition Diffusion Presse}  
\newacro{EIT}{Extreme ultraviolet Imaging Telescope}
\newacro{EPIC}{European participation in Solar-C}
\newacro{ERC}{European Research Council}
\newacro{ESA}{European Space Agency}
\newacro{EST}{European Solar Telescope}
\newacro{EUV}{extreme ultraviolet}
\newacro{FAF}{flaring active-region fibril}
\newacro{FITS}{Flexible Image Transport System}
\newacro{FOV}{field of view}
\newacro{fov}{field of view}
\newacro{FWHM}{full width at half maximum}
\newacro{HAO}{High Altitude Observatory}
\newacro{HD}{hydrodynamics}
\newacro{Hi-C}{High Resolution Coronal Imager Sounding Rocket}
\newacro{HMI}{Helioseismic and Magnetic Imager}
\newacro{IAA}{Instituto de Astrof\'{i}sica de Andaluc\'{i}a}
\newacro{IAC}{Instituto de Astrof\'{i}sica de Canarias}
\newacro{IAS}{Institut d'Astrophysique Spatiale}
\newacro{IBIS}{Interferometric Bi-dimensional Spectrometer}
\newacro{IDL}{Interactive Data Language}
\newacro{IMaX}{Imaging Magnetograph eXperiment}
\newacro{INAF}{Istituto Nazionale di Astrofisica}
\newacro{IB}{IRIS bomb}
\newacro{IR}{infrared}
\newacro{IRIS}{Interface Region Imaging Spectrograph}
\newacro{ISAS}{Institute of Space and Astronautical Science}
\newacro{ISP}{Institute for Solar Physics}
\newacro{ISS}{International Space Station}
\newacro{ISSI}{International Space Science Institute}
\newacro{ITA}{Institute for Theoretical Astrophysics}
\newacro{JAXA}{Japan Aerospace Exploration Agency}
\newacro{KIS}{Kiepenheuer--Institut f\"{u}r Sonnenphysik}
\newacro{KPNO}{Kitt Peak National Observatory}
\newacro{LASP}{Laboratory for Atmospheric and Space Physics}
\newacro{LC}{liquid cristal}
\newacro{LMSAL}{Lockheed Martin Solar and Astrophysics Labratory}
\newacro{LOS}{line of sight}
\newacro{LTE}{local thermodynamic equilibrium}
\newacro{MC}{magnetic concentration}
\newacro{MCAO}{multi-conjugate adaptive optics} 
\newacro{MDI}{Michelson Doppler Imager}
\newacro{ME}{Milne-Eddington} 
\newacro{MHD}{magnetohydrodynamics}
\newacro{MOMFBD}{Multi-Object Multi-Frame Blind Deconvolution}
\newacro{MPE}{Max--Planck--Institut f\"ur extraterrestrische Physik}
\newacro{MPG}{Max--Planck--Gesellschaft}
\newacro{MPS}{Max Planck Institute for Solar System Research}
\newacro{MSSL}{Mullard Space Science Laboratory}
\newacro{MTF}{modulation transfer function}
\newacro{NAOJ}{National Astronomical Observatory of Japan}
\newacro{NASA}{National Aeronautics and Space Administration}
\newacro{NLTE}{non-local thermodynamic equilibrium}
\newacro{NLFFF}{non-linear force-free field}
\newacro{NOAA}{National Oceanic and Atmospheric Administration}
\newacro{non-E}{non-equilibrium}
\newacro{NSO}{National Solar Observatory}
\newacro{NWO}{Netherlands Organisation for Scientific Research}
\newacro{PHE}{propagating heating event}
\newacro{PRD}{partial redistribution}
\newacro{PROBA2}{PRoject for OnBoard Autonomy}
\newacro{PSBE}{post Saha-Boltzmann extinction}
\newacro{PSF}{point spread function}
\newacro{QS}{quiet Sun}
\newacro{QSEB}{quiet-Sun Ellerman-like brightening} 
\newacro{RAL}{Rutherford Appleton Laboratory}
\newacro{RBE}{rapid blue-shifted excursion}
\newacro{R-MHD}{radiation hydrodynamics}
\newacro{rms}{root mean square}
\newacro{RMS}{root mean square}
\newacro{ROB}{Royal Observatory of Belgium}
\newacro{ROI}{region of interest}
\newacro{RRE}{rapid red-shifted excursion}
\newacro{RTE}{radiative transfer equation}
\newacro{SE}{statistical equilibrium}
\newacro{SB}{Saha Boltzmann}
\newacro{SDO}{Solar Dynamics Observatory}
\newacro{SJI}{slit-jaw image}
\newacro{SNR}{signal-to-noise ratio}
\newacro{SO}{Solar Orbiter}
\newacro{SoHO}{Solar and Heliospheric Observatory}
\newacro{SP}{Spectropolarimeter}
\newacro{SST}{Swedish 1-m Solar Telescope}
\newacro{SUMER}{Solar Ultraviolet Measurements of Emitted Radiation}
\newacro{SUFI}{Sunrise Filter Imager}
\newacro{SVD}{singular value decomposition}
\newacro{SVST}{Swedish Vacuum Solar Telescope}
\newacro{THEMIS}{T\'{e}lescope H\'{e}liographique pour l'Etude du 
   Magn\'{e}tisme et des Instabilit\'{e} Solaires}     
\newacro{TR}{transition region}
\newacro{TRACE}{Transition Region and Coronal Explorer}
\newacro{TSI}{total solar irradiance}
\newacro{UT}{Universal Time}
\newacro{UV}{ultraviolet}
\newacro{VAULT}{Very high angular resolution ultraviolet telescope}
\newacro{VIRGO}{Variability of solar IRradiance and Gravity Oscillations}
\newacro{VTT}{Vacuum Tower Telescope}    
\newacro{XRT}{X-Ray Telescope}

\def\acp#1{\pdftooltip{\acs{#1}}{\acl{#1}}} 

\def\nl{,\ } 

\def\LA{Lingezicht Astrophysics\nl 't Oosteneind 9\nl 4158\,CA Deil\nl 
        The Netherlands}

\def\NAOJ{National Astronomical Observatory of Japan\nl
          2-21-1 Osawa, Mitaka\nl Tokyo 181-8588\nl Japan}

\long\def\startignore #1\stopignore{}   

\def\rmit#1{{\it #1}}              
\def\etal{\rmit{et al.}}           
           
\def\ie{\rmit{i.e.,}}              
\def\eg{\rmit{e.g.,}}              
\def\cf{cf.}                       

\def\specchar#1{\uppercase{#1}}    

\def\CaII{\mbox{Ca\,\specchar{ii}}}

\def\HI{\mbox{H\,\specchar{i}}} 
 
\def\Hmin{\hbox{{\rm H}$^{^{_-}}$}}      

\def\Halpha{\mbox{H\hspace{0.1ex}$\alpha$}} 

\def\Lyalpha{\mbox{Ly$\hspace{0.2ex}\alpha$}}

\def\CaIIK{\mbox{Ca\,\specchar{ii}\,\,K}}       
\def\CaIIH{\mbox{Ca\,\specchar{ii}\,\,H}}

\def\HtwoV{\mbox{H$_{2V}$}}

\def\CaIR{\mbox{Ca\,\specchar{ii}\,8542\,\AA}} 

\def\level #1 #2#3#4{$#1 \; ^{#2} \mbox{#3} ^{#4}$}   

\def\rmd{{\rm d}} \def\rmD{{\rm D}} 
\def\rme{{\rm e}}

 \def\rmH{{\rm H}}

 \def\rmK{{\rm K}}

\def\tis{\!=\!}                            
\def\tapprox{\!\approx\!}                  

\def\rmit#1{#1}                 
\def\specchar#1{{\textsc{#1}}}  

\ifnum\markrevised=1
  \long\def\rev#1{{{\bf #1}}}   
  \def\revdel{{{\!\boldmath$\forall$\,}}}  
\else
  \long\def\rev#1{#1}           
  \def\revdel{}                
\fi

\bibnote{2016A&A...590A.124R}{(Pub~I)}  
\def\PubI{\href{http://adsabs.harvard.edu/abs/2016A&A...590A.124R}{Pub~I}}
\bibnote{2016arXiv160901122R}{(Pub~II)}  
\def\PubII{\href{http://adsabs.harvard.edu/abs/2016arXiv160901122R}{Pub~II}} 

\def\linkssx#1#2{\href{http://www.staff.science.uu.nl/~rutte101/rrweb/rjr-pubstuff/lyalma/cutssx.pdf\#page=#1}{#2}}  
\def\linkssxp#1#2{\href{http://www.staff.science.uu.nl/~rutte101/rrweb/rjr-pubstuff/lyalma/cutssx.pdf\#page=#1}{display #1: #2}}  

\begin{document}

\title{Solar H-alpha features with hot onsets} 
\subtitle{III. Long fibrils in Lyman-alpha and with 
ALMA\thanks{
This study is offered as compliment to M.W.M. de Graauw. 
Our ways, objects, instruments and spectral domains parted after the
1970 eclipse but converge here.}}
\author{Robert J. Rutten\inst{1,2}}

\institute{\LA\ \and \NAOJ} 

\date{Received 4 July 2016 /  Accepted 4 October 2016}   

\abstract{In H-alpha most of the solar surface is covered by dense
canopies of long opaque fibrils, but predictions for
quiet-Sun observations with ALMA have ignored this fact. 
Comparison with Ly-alpha suggests that the large opacity of H-alpha
fibrils is caused by hot precursor events.
Application of a recipe that assumes momentary Saha-Boltzmann
extinction during their hot onset to millimeter
wavelengths suggests that ALMA will observe H-alpha-like fibril
canopies, not acoustic shocks underneath, and will yield data
more interesting than if these canopies were transparent.} 


\maketitle

\section{Introduction}\label{sec:introduction}
The Sun consists largely of hydrogen. 
The diagnostics provided by hydrogen are therefore of prime interest
in studies of the solar atmosphere, in particular the chromosphere.    
Here I discuss the principal ones, \Lyalpha\ at 1216\,\AA, \Halpha\ at
6563\,\AA, and the millimeter continua accessible to the Atacama Large
Millimeter/submillimeter Array (ALMA,
\href{http://www.almaobservatory.org}{website}) jointly
by estimating and comparing the visibilities of chromospheric features
in these diverse spectral windows.

The key observation prompting this study is the well-known fact that
in the center of \Halpha\ most of the solar surface is covered by
opaque \rev{slender} fibrils,
\rev{not only for active regions as in} \revdel
Fig.~\ref{fig:halya} below \rev{but also in quieter areas
(\eg\ \linkadspage{2007ASPC..368...27R}{9}{Figs.~6--9} of
\citeads{2007ASPC..368...27R}) 
and even (as so-called mottles) in very quiet areas as} in
\linkadspage{2007ApJ...660L.169R}{2}{Fig.~1} of
\citetads{2007ApJ...660L.169R}, 
except for relatively rare \rev{super-}quiet locations 
(upper-left corner in
\rev{\linkadspage{2007ApJ...660L.169R}{2}{Fig.~1}} of
\citeads{2007ApJ...660L.169R}, 
lower-left corner in \rev{\linkadspage{2007ASPC..368...27R}{13}{Fig.~9}} of
\citeads{2007ASPC..368...27R}). 
\revdel

The question addressed here is how these fibrils will appear in
\acp{ALMA} images. 
My prediction: even more opaque.

The key premise of this study is that the \Halpha\ fibril canopy stems
from propagating heating events (PHE) that belong to a
wide-ranging family
emanating from magnetic concentrations in network and plage.
The most familiar members of this family are off-limb
spicules-II (\citeads{2007PASJ...59S.655D}; 
\citeads{2007Sci...318.1574D}) 
and their on-disk representations as rapid
blue-wing excursions in \Halpha\ (RBE,
\citeads{2008ApJ...679L.167L}; 
\citeads{2009ApJ...705..272R}; 
\citeads{2012ApJ...752..108S}) 
and similar rapid red-wing excursions in \Halpha\ (RRE,
\citeads{2013ApJ...769...44S}), 
although their drivers remain unidentified
(\citeads{2012ApJ...759...18P}). 
These must be more energetic than simple acoustic shocks running up
slanted fluxtubes because the latter produce well-understood dynamic
fibrils around plage and active network
(\citeads{2006ApJ...647L..73H}; 
\citeads{2007ApJ...655..624D}) 
and similar but shorter dynamic fibrils in active regions
(\citeads{2013ApJ...776...56R}). 

I postulate below that there must be more horizontal \acp{PHE}s that
produce the observed ubiquitous long \Halpha\ fibrils. 
My working hypothesis is that the latter are post-\acp{PHE} contrails,
comparable to the contrails drawn by passing aircraft on our sky.

The second study in this series describes an example of such
a contrail: a large dark \Halpha\ fibril marking the earlier passage
of a spectacular \acp{PHE} 
(\rev{\citeads{2016arXiv160907616R}}, 
henceforth Pub~II). 
The \acp{PHE} shared properties with more ordinary \acp{RBE}s but it
was unusually large and energetic and was launched in a more
horizontal direction.
Minutes later the large \Halpha\ fibril appeared as if the \acp{PHE}
trajectory became outlined with a fat marker pen. 

Such delayed \Halpha\ contrail formation stems from the enormous
relative abundance of hydrogen and the large excitation energy of its
first excited level ($n \tis 2$ at 10.2~eV).
These properties produce particular formation characteristics that
affect all three spectral windows: slow non-equilibrium
ionization/recombination balancing at low temperature, Saha-Boltzmann
or near-Saha-Boltzmann high-level and ion populations at high
temperature and density, large extinction at high temperature that
remains large in cool \acp{PHE} aftermaths.

These extraordinary characteristics are elaborated in
Section~\ref{sec:visibilities} with reference to and in sequel to the
first study in this series
(\citeads{2016A&A...590A.124R}, 
henceforth Pub~I). 
It formulated a hot-onset recipe to understand the diverse
visibilities of Ellerman bombs (photospheric reconnection events;
review by \citeads{2013JPhCS.440a2007R}). 
In brief, the recipe is to evaluate the extinction of \Halpha\ at the
onset of a hot event assuming Saha-Boltzmann population of the
hydrogen $n \tis 2$ level and to sustain this high value during
subsequent cooling, also more widely around it.
I suggest here that this post-Saha-Boltzmann-extinction (PSBE) recipe
also applies to the long fibrils constituting the \Halpha\ canopy.

Section~\ref{sec:Lya-Ha} compares solar \Halpha\ and \Lyalpha\ images
as major motivation to invoke \acp{PHE}'s as fibril generator.  
Additional evidence comes from the paucity of internetwork shock
scenes in \Halpha\ and fibril incongruity between \Halpha\ and \CaIR\
(Sect.~\ref{sec:discussion}).

Section~\ref{sec:discussion} also discusses how to detect these
\acp{PHE}s, why numerical simulations so far fail to reproduce the
actual \Halpha\ chromosphere, and what fibril widths and
temperatures \acp{ALMA} will measure.

The conclusion (Sect.~\ref{sec:conclusion}) summarizes this study by
predicting that in \acp{ALMA} images the Sun will appear as the
fibrilar \Halpha\ chromosphere, largely obscuring the
shock-interference internetwork scenes that have been extensively
foretold in quiet-Sun predictions for \acp{ALMA} based on
numerical simulations (details, reviews and references in Wedemeyer
\etal\ \citeyearads{2016SSRv..200....1W}, 
\citeyearads{2016arXiv160100587W}). 

\section{Extinction in and and after hot precursors}     
\label{sec:visibilities}

\subsection{PSBE recipe}
\PubI\ based the \acp{PSBE} recipe for hot-onset features on the
one-dimensional plane-parallel hydrostatic model atmosphere of
\citetads{2008ApJS..175..229A} 
and the two-dimensional non-equilibrium \acp{MHD} simulation of
\citetads{2007A&A...473..625L}, 
regarding these as the atmospheres of hypothetical stars called ALC7
and HION.
They share properties with the Sun: in the case of ALC7 the emergent
disk-center average spectrum, in the case of HION the presence of
granulation, sound waves, acoustic internetwork shocks, network-like
magnetic fields, and even dynamic fibrils. 
In my opinion HION comes much closer to the actual Sun even while
being only two-dimensional, whereas ALC7 is a superb didactic star of
which the detailed spectrum is easily synthesized with the splendid
virtue of being fully understandable\footnote{For students of
\wlRRcourses{my courses}.
Recently I added ALC7 line formation diagrams to 
\wlRRssx{my example displays} when invited to teach at the \acp{NAOJ}.
Some are referenced here as as \linkssx{1}{online material}.} 

The ALC7 and HION properties inspiring the \acp{PSBE} recipe for
\Halpha\ features are:
\begin{itemize} \vspace{-0.5ex} \itemsep=1ex

\item wherever the \Lyalpha\ source function $S$ equals the Planck
function $B(T)$ for temperature $T$ and hydrogen is predominantly
neutral the extinction coefficient of \Halpha\ is given by the
Saha-Boltzmann (henceforth SB) value because the \HI\ ground level is
then the population reservoir with \acp{NLTE} population
departure coefficient $b_1 \tapprox 1$ and $S \tapprox (b_2/b_1)\,B$
(\rev{\linkadspage{2016A&A...590A.124R}{2}{Sect.\,2.1} of} \PubI);

\item although \Lyalpha\ is the most scattering line in the solar
spectrum it thermalizes in the ALC7 chromosphere.
The thermalization length for its Doppler core in optical path units
is $\Lambda^\rmD_\tau = \sqrt{\pi}/\varepsilon$, with $\varepsilon$
the collisional destruction probability per line photon extinction
(\citeads{1965SAOSR.174..101A}). 
This relation holds for an isothermal constant-$\varepsilon$
atmosphere, but it is a good approximation for the ALC7 chromosphere
which is nearly isothermal and has near-constant
$\varepsilon \tapprox 10^{-6}$ because increasing hydrogen ionization
compensates the density decrease (\linkssxp{2}{ALC7 model}).
The very small $\varepsilon$ implies thermalization over a million
photon path lengths in ALC7, the most of all chromospheric lines
(\linkssxp{18}{\Lyalpha}).
However, even such huge scattering extent remains geometrically small
because the \Lyalpha\ line-center extinction coefficient $\alpha$ is
also huge and the corresponding photon mean free path $1/\alpha$ very
small, so that $\Lambda^\rmD_x = \sqrt{\pi}/(\alpha\,\varepsilon)$
increases from only 1\,km at the bottom of the ALC7 chromosphere
($h\tis800$\,km) to 100\,km at $h \tis 1400$\,km where \Halpha\
escapes (\linkssxp{19}{\Halpha}).
This estimate is for the Doppler core only.
At larger density the Stark wings of \Lyalpha\ become more important,
including partially coherent scattering, and increase the actual
thermalization length {$\Lambda_x$} over which the profile-summed and
angle-averaged radiation $\overline{J}$ saturates to the local $B$.  
The wing-escape increase is beyond analytic formulation; numerical
tests with the RH code of
\citetads{2001ApJ...557..389U} 
showed that $\Lambda_x \tis 10-100$\,km in the lower ALC7
chromosphere.
This relatively small range implies that the \Lyalpha\ radiation
remains locked-in within it and thermalizes to the local temperature
where \Halpha\ escapes, producing $S\tapprox B$ equality in \Lyalpha\
and \acp{SB} extinction for \Halpha\
(\rev{\linkadspage{2016A&A...590A.124R}{3}{Fig.~1}} of \PubI;
\linkssxp{19}{\Halpha});

\item shocks in the HION atmosphere have temperatures of order 7000\,K
and electron densities around 10$^{11}$\,cm$^{-3}$, just as the ALC7
chromosphere which has $T \tapprox 6700$\,K and
$N_\rme \tapprox 10^{11}$\,cm$^{-3}$, both nearly constant
(\linkssxp{2}{ALC7 model}).
This equality is not surprising because the emphasis in the ALC7
best-fit construction was on the quiet-Sun ultraviolet spectrum in
which non-linear Wien $\rmd B / \rmd T$ sensitivity gives larger
weight to higher temperature
(\citeads{1994chdy.conf...47C}) 
while in quiet areas acoustic shocks dominate the internetwork
sub-canopy domain (review in
\citeads{1995ESASP.376a.151R}) 
so that ultraviolet spectrum fitting replicates the shock temperature
spikes
(\citeads{1995ApJ...440L..29C}); 

\item also within the HION shocks the \HI\ $n \tis 2$
population saturated to the \acp{SB} value (bottom panels of
\rev{\linkadspage{2007A&A...473..625L}{5}{Fig.~2}}
of \citeads{2007A&A...473..625L}). 
\revdel
In this simulation the net photon rates in Lyman transitions were put
to zero for tractability following
\citetads{2002ApJ...572..626C}. 
This simplification produced high-lying green fibril-like arches in
the last panel of \rev{\linkadspage{2007A&A...473..625L}{4}{Fig.~1}}
of \citetads{2007A&A...473..625L} 
which are artifacts, but it does not affect lower layers
where radiative Lyman balancing is a good approximation \rev{as
shown for 1D static models by} \citetads{1981ApJS...45..635V} 
\rev{and for HION-like shocks by
\citetads{2002ApJ...572..626C}}; 

\item in the HION shocks hydrogen reaches about 10\% ionization with
$b_c \tapprox 0.1$; they would reach full ionization if \acp{LTE} were
valid (thin curves in the second row of
\rev{\linkadspage{2007A&A...473..625L}{5}{Fig.~2}} of
\citeads{2007A&A...473..625L}). 
Similar \acp{NLTE} underionization occurs through the ALC7
chromosphere (\linkssxp{4}{ALC7 hydrogen}).
It does not represent a transition from \acp{SB} to
coronal-equilibrium (CE) partitioning, which in the ALC7 atmosphere
occurs at much lower density in the transition region.
Instead, it follows the Balmer continuum which is the main hydrogen
ionization agent in the chromosphere (\eg\
\citeads{1981ApJS...45..635V}; 
\citeads{2002ApJ...572..626C}) 
and has constant radiation temperature near 5250\,K as a scattering
average of photons created in the granulation (\linkssxp{4}{ALC7
hydrogen}).
In the HION and ALC7 atmospheres the top of the hydrogen atom behaves
as a 3.4\,eV alkali atom with $n \tis 2$ as ground level having
\Lyalpha-defined constraining population and with the continuum
population at an offset from it that is given by balancing the
$B_{\rm Ba cont}(5250\,\rmK)/B_{\rm Ba cont}(T)$ \acp{NLTE} ionization
driving with photon losses in Balmer and higher lines that govern
\acp{NLTE} recombination
(\rev{\linkadspage{1994IAUS..154..309R}{6}{Fig.~3}} of
\citeads{1994IAUS..154..309R}); 

\item in the HION atmosphere drastic cooling occurs after a shock has
passed. 
However, in this non-equilibrium (non-E) star the \HI\ $n \tis 2$
population \rev{did} not adapt instantaneously.
\revdel Collisional bound-bound balancing has Boltzmann
temperature sensitivity \rev{through the collisional excitation rate
$n_1\,C_{12}$} 
(\rev{\linkadspage{2003rtsa.book.....R}{70}{p.~50--51}} of
\citeads{2003rtsa.book.....R}) 
making the 10.2\,eV \Lyalpha\ jump too large for fast thermal
balancing at low temperature, as demonstrated by
\citetads{2002ApJ...572..626C}. 
\rev{In shocks the collisional balancing is fast so that the
\Lyalpha\ radiation and corresponding $n \tis 2$ population reach the
high \acp{LTE} values, but they then hang} near \rev{these
while the gas cools} \revdel until the next shock passes
(bottom panels of
\rev{\linkadspage{2007A&A...473..625L}{5}{Fig.~2}} of
\citeads{2007A&A...473..625L}). 
The Balmer continuum and lines couple the ionization degree to this
retarded \acp{non-E} $n \tis 2$ behavior without further
retardation, initially giving $b_{\rm cont}/b_2 \tapprox 0.1$ and
then reversing to $b_{\rm cont}/b_2 \!\gg\! 1$ while the gas cools
well below 5250\,K.
The bottom row of
\rev{\linkadspage{2007A&A...473..625L}{4}{Fig.~1}} of
\citetads{2007A&A...473..625L} 
shows that the post-shock HION clouds so reach huge overpopulations:
up to $b_2 \tis 10^{12}$ and $b_{\rm cont} \tis 10^{15}$;

\item hot features embedded in cool gas irradiate their surroundings
in \Lyalpha\ over hundreds of kilometers due to smaller $\varepsilon$
in the cool gas.
This radiation boosts $J$ and therefore $S$ in \Lyalpha\ and with it
the \Halpha\ extinction across such scattering extents towards the
high-temperature values within the feature
(\rev{\linkadspage{2016A&A...590A.124R}{5}{Fig.~3}} of \PubI;
\linkssxp{19}{\Halpha}; \linkssxp{26}{aureole boosting}).
A momentary heating event has similar boost-spreading as long as
hydrogen remains partially neutral. 
This then lasts minutes after its occurrence.

\end{itemize}

\noindent
With these, the \acp{PSBE} recipe for \Halpha\ features became: (1)
evaluate the \Halpha\ extinction coefficient during the hot onset of
a dynamic feature by assuming the \acp{SB} value, (2) use the
resulting large population also for cooler surrounding gas in reach of
scattering \Lyalpha\ radiation, and (3) maintain this large population
subsequently during cooling aftermaths. 
\PubI\ so explained the diverse and even discordant visibilities
of Ellerman bombs that cannot be reproduced with static equilibrium
models.

A comment concerning \acp{ALMA}: \Halpha-like memory of hot instances
in the recent past holds also for hydrogen free-free continua
since these are similarly extinction-boosted by retarded hydrogen
recombination in cool post-hot episodes, \ie\ cool gas where a
heating event passed shortly before. 
Below I concentrate on long \Halpha\ fibrils and my
interpretation of these as post-hot contrails, but similar
\acp{non-E} boosting must take place in decidedly post-hot
phenomena as spicules-II and coronal rain to which I return in the
conclusion.

A comment concerning \Lyalpha: its extinction does not suffer from
retarded collisional balancing after HION shocks because at
10\% ionization the \HI\ ground level remains the population reservoir
in these.
However, it might when recombination follows on hotter precursor
events that ionize hydrogen more completely. 

\begin{figure*}
  \includegraphics[width=\textwidth]{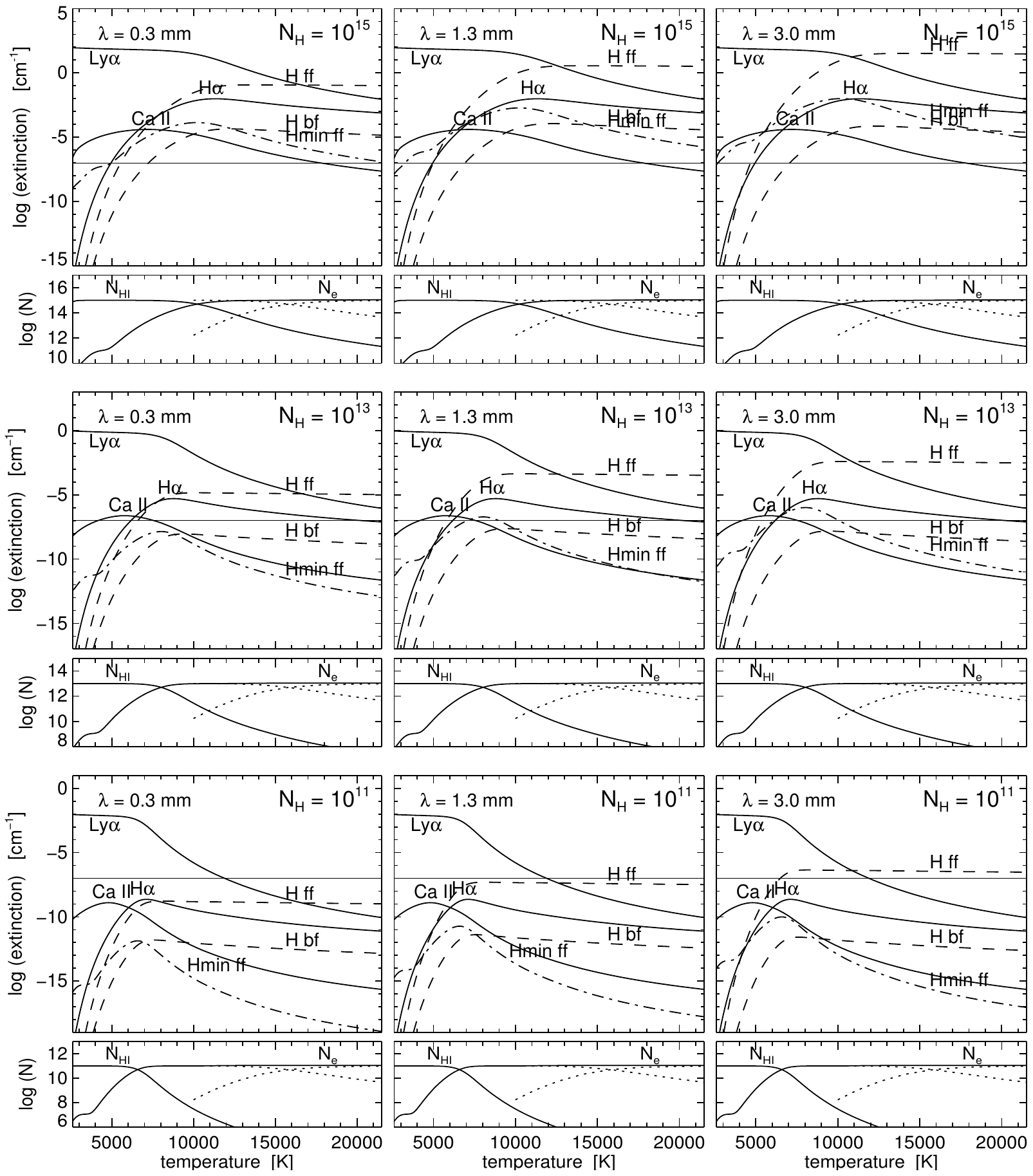}
  \caption[]{\label{fig:extalma} 
  Saha-Boltzmann (SB) line extinction coefficient against
  temperature at the centers of \Lyalpha, \Halpha\ and \CaIR\ (solid)
  and \acp{SB} continuous extinction coefficient of the \HI\
  free-free and bound-free contributions (dashed) and the \Hmin\
  free-free contribution (dot-dashed) at three \acp{ALMA} wavelengths
  (from left to right 0.35, 1.3 and 3.0\,mm), for gas of solar
  composition with different total hydrogen densities (from top to
  bottom $N_\rmH\tis10^{15}, 10^{13}$ and $10^{11}$~cm$^{-3}$,
  corresponding to the bottom, middle and top of the ALC7
  chromosphere).
  The horizontal line at $y \tis -7$ specifies optical thickness unity
  for a line of sight through a 100\,km wide slab.
  The small panels underneath each extinction graph
  show the competing neutral hydrogen and electron densities
  (solid, particles\,cm$^{-3}$). 
  The overlaid dotted curves starting at 10\,000~K are
  coronal-equilibrium (CE) neutral hydrogen and proton densities from
  \wlchianti{CHIANTI}.
  The density scales have the same logarithmic unit (dex) size
  as the extinction scales to enable slope comparisons.   
  The $y$-axis scales shift between rows for better curve visibility,
  \vspace{2ex}\mbox{}} 
\end{figure*}

\subsection{Saha-Boltzmann extinction}
Figure~\ref{fig:extalma} shows monochromatic extinction
coefficients (large panel per pair) and hydrogen and electron
densities (small panel per pair) in a parcel of solar gas with
given total hydrogen density as function of temperature, assuming
\acp{SB} population partitioning. 
It was made, similarly to
\rev{\linkadspage{2016A&A...590A.124R}{9}{Fig.~5}} in \PubI, with
\acp{IDL} programs (\wlRRidl{on my website}) based on \acp{LTE}
programs from a
\href{https://github.com/aasensio/lte/tree/master/idl}{github
repository} of A.~Asensio Ramos that were written by J.~S\'anchez
Almeida (\citeyearads{1992SoPh..137....1S}, 
\citeyearads{1997ApJ...491..993S}) 
and were partly based on
\citetads{1974SoPh...35...11W} 
following \citet{Mihalas1967}. 
I renewed and extended these programs with data and routines in
the \wlsolarsoft{SolarSoft} \wlchianti{CHIANTI package} (\eg\
\citeads{1997A&AS..125..149D}, 
\citeads{2013ApJ...763...86L}). 
The dotted comparison curves for coronal equilibrium (CE) in the
small density panels are directly from \wlchianti{CHIANTI}. 

The horizontal line at $y \tis -7$ in each extinction panel
specifies optical thickness unity for a line of sight crossing a
100-km wide slab.
Extinction above this line implies optically thick sampling of
a feature of this size. 

The extinction curve patterns with their initial rises (except
\Lyalpha), peaks, and steep to slow declines for increasing
temperature are primarily set by hydrogen ionization, in which the
cross-over from atoms to protons shifts left and steepens from row to
row.
The extinction patterns change correspondingly but retain their
qualitative offset and cross-over behavior.
The main changes are the decreases $\propto\!N_\rmH$ or
$\propto\!N_\rmH\,N_\rme$ from row to row, most clearly exhibited by
the curve separations from the $y \tis -7$ thickness indicator line,
and the \HI\ free-free extinction increase $\propto\!\lambda^2$ along
rows (Eq.~(5.19a) of \citeads{1986rpa..book.....R}; 
\rev{\linkadspage{2003rtsa.book.....R}{47}{Eq.~(2.79)}} 
of \citeads{2003rtsa.book.....R}). 

The $\alpha_{\rm ff}\!\propto\!\lambda^2$ increase shifts
the $\tau \tis 1$ height of radiation escape across the ALC7
chromosphere. 
At 0.35\,mm it lies at its bottom, at 3\,mm at its top.
Because the rows of Fig.~\ref{fig:extalma} cover this
range in gas density and the ALC7 temperature is nearly constant at
6700\,K, the corresponding \acp{SB} extinction coefficients can be
read off as 6700\,K samplings of the three diagonal (first-to-last)
panels of Fig.~\ref{fig:extalma}.
The HION shocks have total hydrogen density
$N_\rmH \approx 10^{14}$, between the first two rows.

The actual hydrogen ionization follows cooler Balmer radiation as
noted above, but this \acp{NLTE} departure produces only slight
desteepening of the steep increases in the \HI\,ff curves around the
$T\tis5250$\,K pivot.
For ALC7 it represents only a one dex tilt correction of
the 10~dex increase.
For HION it amounts to correction by $+3$\,dex at 3000\,K in the
shock aftermaths diminishing to $-1$\,dex at 7000\,K in the shocks
themselves, still relatively small.

I added the dotted \acp{CE} curves to the density panels to
indicate hydrogen ionization at lower gas densities.
They are unrealistic at chromospheric densities
but they do show the trend because from very high to very low
density the truth shifts from \acp{SB} to \acp{CE}.
The invariant \acp{CE} curve cross-over at 16\,000\,K implies
that at decreasing density the leftward shift and
steepening of the \acp{SB} hydrogen cross-overs get compensated.

I now discuss the spectral features in Fig.~\ref{fig:extalma} one by
one.

\Lyalpha\ is of course the extinction champion at low temperature.
Since its lower level then contains virtually all hydrogen \acp{SB}
extinction is guaranteed.
At higher temperatures its extinction-coefficient decline follows the
neutral-hydrogen density decline.
In \acp{CE} this is much less steep than for \acp{SB}, predicting
higher-temperature \Lyalpha\ presence for lower densities.
A solar feature needs to be only kilometer-size to get optically thick
in this line.
Only above 40\,000\,K \Halpha\ obtains larger \acp{SB} extinction from
the $g_2/g_1 \tis 8$ statistical-weight ratio.
At lower temperature anything opaque in \Halpha\ is necessarily much
more opaque in \Lyalpha. 

For \Halpha\ the \acp{SB} extinction assumption is correct when
hydrogen ionization at high density increases $\varepsilon$ in
\Lyalpha\ sufficiently within opaque features. 
As noted above $b_2 \tapprox b_1 \tapprox 1$ indeed holds throughout
the ALC7 chromosphere and also within the HION shocks. 
At higher temperature the corresponding increase in
$\varepsilon$ produces \Lyalpha\ thermalization within yet smaller
features.

For \CaIR\ the assumption of \acp{SB} extinction is also reasonable,
actually an underestimate where photon losses in the \CaII\ infrared
lines cause \acp{NLTE} overpopulation of their lower levels
(\rev{\linkadspage{2016A&A...590A.124R}{3}{Fig.~1}} of \PubI,
\linkssxp{15}{\CaIR}).
Observationally, this line shows fibrils comparably to \Halpha\ near
network but not further out from network, an important incongruity to
which I return in Sect.~\ref{sec:discussion}.

\HI~free-free extinction behaves remarkably similar to \Halpha\
in its steep initial increases.
Beyond these this contribution dominates the \acp{ALMA} continua
and grows steeply $\propto\!\lambda^2$ while
decreasing slowly $\propto\!T^{-3/2}$ across these
parameter ranges.
The \HI\,ff\,/\,\Halpha\ extinction coefficient ratio
increases with both to very large values.
Anything opaque in \Halpha\ will be at least similarly opaque at
mm wavelengths. 
Anything hot and opaque in \Halpha\ will be much more opaque at
the longer \acp{ALMA} wavelengths, even exceeding \Lyalpha\ above
12\,000\,K.  

\HI\ bound-free extinction is less important than \HI\ free-free
extinction in all panels.

H-minus free-free extinction is only important below about 5000\,K
where the small plateau in the first panel and in the electron
density curves illustrates that metal ionization rather than hydrogen
ionization governs the optical and infrared continua from the
photosphere. 
The upper photosphere is fully transparent in \Halpha\ but not at
mm wavelengths.  

The extinction coefficients of Thomson and Rayleigh scattering
were also evaluated, but they are not shown because they are not
competitive in this parameter domain.

\subsection{Non-equilibrium extinction}
The curves in Fig.~\ref{fig:extalma} are for instantaneous \acp{LTE}
but permit interpretation of the \acp{non-E} results of the
HION simulation. 
While \Lyalpha\ thermalizes within the HION shocks giving \acp{SB}
extinction to \Halpha, during the post-shock cooling the gas
temperature follows the very steep Boltzmann slopes left of the
\Halpha\ peaks in Fig.~\ref{fig:extalma} and so the \Halpha\
extinction would get far smaller, over 10 orders of magnitude, if
\acp{SB} remained valid.
However, this is not the case: retarded \Lyalpha\ settling maintains
the high $n \tis 2$ population reached in the shocks and so produces
the huge $b_2$ values in the last panels of
\rev{\linkadspage{2007A&A...473..625L}{4}{Fig.~1}} and
\rev{\linkadspage{2007A&A...473..625L}{5}{Fig.~2}} of
\citetads{2007A&A...473..625L}. 
These are called ``\acp{NLTE} overpopulations'' but a better name
would be ``\acp{SB} underestimates''. 
In the HION shock aftermaths the temperature drops typically from
7000\,K to 3000\,K, indeed corresponding to 10~dex \acp{SB}
underestimate for $n\tis2$ due to the very steep Wien
$\rmd B / \rmd T$ at \Lyalpha.  
The much larger actual population therefore gives \Halpha\ \acp{PSBE}
formation in HION post-shock cooling clouds.
Figure~\ref{fig:extalma} suggests that such time-lagged \acp{PSBE}
boosting may likewise occur in \HI\,ff continua, with
large boosts at longer wavelengths.

\begin{figure*}
  \centerline{\includegraphics[width=\textwidth]{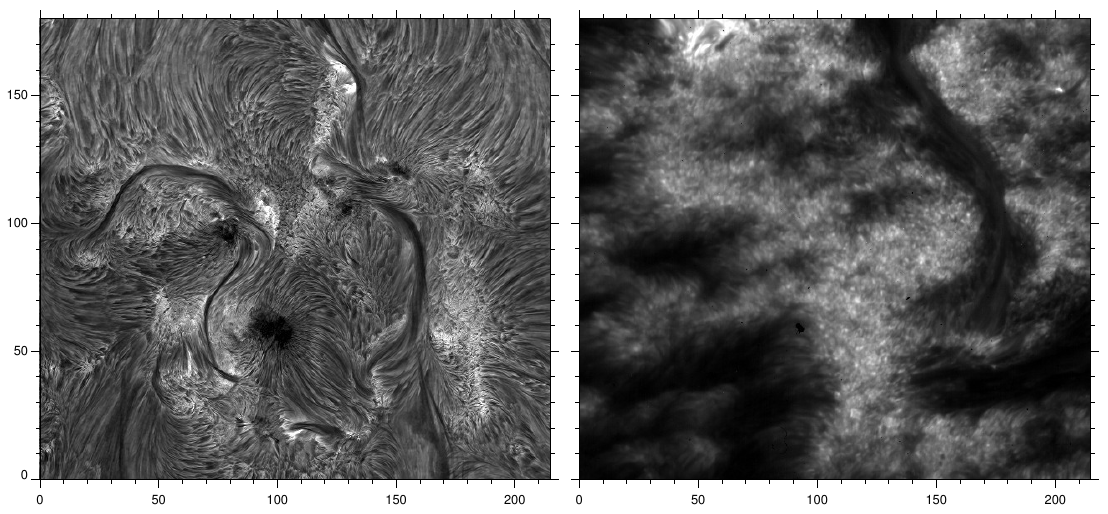}}
    \caption[]{\label{fig:halya} 
    Comparison of solar scenes in \Halpha\ and \Lyalpha.
    Scales in arcsec.
    The images have identical field size but sample the Sun at
    different times (July 8, 2005 and June 14, 2002) and viewing
    angles ($\mu\tis 0.98$ and $\mu\tis 0.68$ with the limb to the top).
    The first is an \Halpha\ line-center mosaic constructed by
    P.~S\"utterlin from nine images he took with the \acl{DOT} (DOT,
    \citeads{2004A&A...413.1183R}). 
    The second is part of the 13th wide-band \Lyalpha\ image from
    \acp{VAULT}-II (\citeads{2010SoPh..261...53V}). 
    The \acp{DOT} pixels were 0.071\,arcsec, the \acp{VAULT} pixels
    0.124\,arcsec. 
    The claimed resolution is 0.3\,arcsec\ for both, but zoom-in per
    pdf viewer suggests that the \acp{DOT} image comes closer to this
    value than the \acp{VAULT} image.
    }
\end{figure*}

\subsection{Source functions and intensities}
\acp{ALMA} is correctly advertised as ``linear thermometer'' for
optically thick structures in the solar atmosphere. 
The mm-wavelength $S \tis B \tis (2ck/\lambda^4)\,T$ equality indeed
holds wherever the kinetic Maxwell distribution holds since free-free
transitions are always collisional, free-free extinction
dominates over the whole parameter domain of Fig.~\ref{fig:extalma},
and the Rayleigh-Jeans limit applies very well. 
The advertising might add that the \acp{LTE} nature of
the source function guarantees sharp detail rendition without
blurring from scattering within the solar atmosphere, \ie\ that
the image resolution is limited only by \acp{ALMA} itself. 

The numerical thermometer demonstration in
\rev{\linkadspage{2016Msngr.163...15W}{4}{Fig.~4}} of
\citetads{2016Msngr.163...15W}, 
copied from \rev{\linkadspage{2007A&A...471..977W}{4}{Fig.~1}} of
\citetads{2007A&A...471..977W}, 
is therefore just an elaborate demonstration of the simple
Eddington-Barbier approximation for the observed intensity
$I \tapprox S(\tau_\mu \tis 1) \tis B(\tau_\mu \tis
1)\!\propto\!T(\tau_\mu \tis 1)$
with $\tau_\mu$ the summed extinction along the line of sight.
A much earlier and more elegant demonstration was the careful labeling
$S\!\tis\!B$ added by E.H.~Avrett to all continuum-formation
displays for wavelengths above 1.6\,$\mu$m in the wonderful 
\rev{\linkadspage{1981ApJS...45..635V}{33}{Fig.~36}} of
\citetads{1981ApJS...45..635V}. 

The issue in interpreting solar observations from \acp{ALMA} is
therefore not a source function issue but an extinction issue.
What is the thermometer sampling? 
More precisely: is the extinction defining the $\tau_\mu \tis 1$ depth
in optically thick features given by the present or by the past? 
In the first case the opacity of features in \acp{ALMA}
images is regulated instantaneously and classical hydrogen ionization
modeling assuming instantaneous statistical equilibrium suffices. 
In the second case the feature opacities depend on what happened
before and the modeling must be \acp{non-E} including gas
histories 
\rev{even in chromospheric conditions 
(\cf\ 
\citeads{1976A&A....47...65K}; 
\citeads{1976ApJ...205..499K}; 
\citeads{1979SoPh...64...57J}; 
\citeads{1979SoPh...61..389P}; 
\citeads{1980A&A....87..229K}; 
\citeads{2002ApJ...572..626C}; 
\citeads{2006ASPC..354..306L}; 
\citeads{2007A&A...471..977W}; 
\citeads{2007A&A...473..625L}).} 


In contrast, \Lyalpha\ has instantaneous extinction
wherever hydrogen is mostly neutral but it is the quintessential
two-level scattering line in the solar spectrum with very small
$\varepsilon$.
Any sizable feature will harbor the well-known scattering decline from
$S \tapprox B$ inside to very low
$S \tapprox \sqrt{\varepsilon}B$ at its surface
(\rev{\linkadspage{1965SAOSR.174..101A}{12}{Fig.~1ff of}} 
\citeads{1965SAOSR.174..101A}; 
\citeads{1970stat.book.....M}; 
\rev{\linkadspage{2003rtsa.book.....R}{112}{Sect.~4.3 of}}
\citeads{2003rtsa.book.....R}), 
unless it has steeply outward increasing temperature near its
surface along the line of sight (as when that is the transition
to the corona). 

\Halpha\ has been regarded as ``photo-electrically controlled'' since
\citetads{1957ApJ...125..260T} 
and
\citetads{1959ApJ...129..401J}, 
meaning that the multi-level detour terms specified by $\eta$ in the
general breakdown of the line source function
$S^l \tis (1-\varepsilon-\eta)\overline{J} + \varepsilon B + \eta
S^\rmD$
dominate over the resonance scattering terms set by $\varepsilon$.  
However, in standard model atmospheres such as ALC7 this is not
correct for the chromospheric layers: there even \Halpha\ is primarily
a two-level scattering line with $S^l \tapprox \overline{J}$
(\rev{\linkadspage{2012A&A...540A..86R}{9}{Fig.~12 of}}
\citeads{2012A&A...540A..86R}; 
\linkssxp{19}{\Halpha}).

The upshot is that for optically thick features darkness (low
brightness temperature) in \acp{ALMA} images indeed means low gas
temperature around the Eddington-Barbier depth $\tau_\mu \tis 1$, with
sharp rendering of detail since there is no scattering.
In contrast, at the centers of \Lyalpha\ and \Halpha\ darkness
generally does not suggest low temperature but large opacity
bringing $\tau_\mu \tis 1$ further out along the
scattering decline, with blurring from resonance scattering.

Similarly, at mm wavelengths bright optically thick features directly
imply correspondingly high temperature around $\tau_\mu \tis 1$ with
sharp rendering, but in scattering line cores
brightness stems from much deeper-sited heating or a very steep
negative $T(\tau_\mu)$ gradient or domination by recombination
detours, again with blurring from scattering.

\subsection{Conclusion of this section}
Hot precursor events such as the shocks in the HION atmosphere have
large \acp{PSBE} opacity in \Halpha\ during the subsequent cooling
phase. 
Similar or larger post-hot extinction is expected at mm
wavelengths.
For hotter onset features much more \acp{PSBE} opacity is
expected, with larger boosts at longer wavelengths. 
Scattering blurs such events in \Lyalpha\ and \Halpha\ but not in mm
continua.

\section{Ly\,$\alpha$ versus H\,$\alpha$}    \label{sec:Lya-Ha}
Figure~\ref{fig:halya} compares solar active-region scenes in the core
of \Halpha\ and wide-band \Lyalpha. 
Regretfully, they are far from simultaneous or cospatial, but they do
exhibit comparable solar scenes.

Why they differ so much has long puzzled me.
Anything visible in \Halpha\ has much larger opacity in \Lyalpha\ so
that any scene observed in \Halpha\ should be exaggerated in \Lyalpha. 
However, the scenes in Fig.~\ref{fig:halya}  are very dissimilar.

Both lines are strongly resonant scattering.
For \Halpha\ it implies lower intensity at larger opacity from
sampling the outward scattering decline further out.
For \Lyalpha\ larger opacity likewise produces a deeper
self-absorption dip at line center, but the \acp{VAULT} images sum the
full profile and are dominated by the profile peak heights and widths.
The peaks scatter independently due to partial redistribution and
increase at higher temperature. 
One so expects to see bright \Lyalpha\ features from hot
sheaths around cooler structures that show up relatively dark in
\Halpha.
In particular, one would expect to recognize long dark
\Halpha\ fibrils as extended filamentary bright transition-region
sheets along them
(\citeads{2007ASPC..368...27R}). 

This expectation does hold for dynamic fibrils in the \acp{VAULT}-II
near-limb images (\citeads{2009A&A...499..917K}), 
but the \acp{VAULT}-II disk image in Fig.~\ref{fig:halya} does not show long
fibrils.
The long active-region filaments present in both images appear
similar, by chance even in shape,
except for wider width in \Lyalpha\ that one indeed expects from much
larger opacity at given neutral hydrogen density.
The ``mossy'' activity areas in \Halpha\ harboring bright grains also
seem to have similar counterparts in \Lyalpha, but such areas are much
more wide-spread in \Lyalpha\ whereas the \Lyalpha\ scene lacks the
domination by long fibrils observed in \Halpha\
everywhere around the mossy areas. 
There appears to be no bright transition-region-sheet mapping of
long \Halpha\ fibrils in \Lyalpha, contrary to my
earlier expectation.

The \Lyalpha\ scene does contain grayish fibrilar
features pointing away from active areas, \eg\ spreading left from the
active region in the lower-left corner of Fig.~\ref{fig:halya} above
$x \tis 100$. 
Above $x \tis 50$ there are fans of such features diverging leftward
from more concentrated gray patches presumably at active
network.
Such fibrilar features were called ``loop-like structures'' by
\citetads{2007ApJ...664.1214P} 
and ``threads'' by \eg\
\citetads{2008ApJ...687.1388J}; 
the latter type of fans were called ``comet heads'' by
\citetads{2008ApJ...687.1388J}. 
These \Lyalpha\ fibrils measure about 10\,arcsec, much shorter than
long \Halpha\ fibrils.

In summary, while the \Halpha\ internetwork scene is dominated
by long fibrils, \Lyalpha\ shows only short fibrils jutting out
from activity.
I can only reconcile this striking difference by
postulating that \Lyalpha\ primarily shows hot events while
\Halpha\ fibrils show subsequent cooler aftermaths. 

I therefore suggest that bright \Lyalpha\ grains represent
initial \acp{PHE}s with steep source function increases and
with more horizontal field-aligned launching at the edges of activity
areas following magnetic canopy expansions over adjacent internetwork.
\revdel 
Indeed, when blinking successive co-aligned \acp{VAULT}-II images one
observes substantial proper motion for some. 
Note that \Lyalpha\ scattering gives them extents of order 0.5\,arcsec
even if the actual \acp{PHE} was smaller.

Figure~\ref{fig:halya} then suggests that the more horizontal
\acp{PHE}s leave cooling gas producing \Halpha\ fibrils as in
the example of \PubII.
Such gas necessarily has much larger opacity in \Lyalpha\ than
in \Halpha, but with much smaller opacity contrasts.
In \Halpha\ adjacent fibrils sample different histories,
mutually out of phase with each coming down the steep decline in
Fig.~\ref{fig:extalma} to lower temperature at a few minutes of
retardation after its individual precursor event.
The gas densities in pressure-equilibriated cooling clouds may not
differ much, but their histories and the resulting \Halpha\ opacities that
define their brightness contrasts while they remain optically thick
can differ very much from one to another.
In contrast, wherever hydrogen is predominantly neutral the \Lyalpha\
opacity varies only with the local gas density and hence differs much
less between adjacent fibrils. 

The \Lyalpha\ source function follows the \Lyalpha\ radiation
$J$.
Even while this remains near its high precursor value from being
boxed-in where \Halpha\ escapes inside \Lyalpha-thick cooling clouds,
it drops very much from scattering photon-loss escape towards the
\Lyalpha\ $\tau_\mu\tis 1$ cloud surface which is very much further
out in optical depth (Fig.~\ref{fig:extalma}).
In addition, the large Wien $\rmd B/\rmd T$ non-linearity at
\Lyalpha\ makes such surfaces very dark, resulting in underexposure
unless bright areas are severely overexposed.  

I suggest that these properties together explain the observed
feature-less dark internetwork regions as in the lower-left
corner of the \Lyalpha\ image in Fig.~\ref{fig:halya}.

\Lyalpha\ does have adjacent-fibril opacity contrasts during the
initial cooling phases while hydrogen recombines from full ionization
if that was reached in the precursor \acp{PHE} (as in the
example of \PubII).
I suggest that such contrasts together with high initial temperatures
produce the observed short \Lyalpha\ fibrils including ``comet
heads''.

Obviously one desires \Halpha--\Lyalpha\ comparisons as in
Fig.~\ref{fig:halya} but co-temporal, co-spatial and as a high-cadence
time sequence, at the same resolution or yet better.
Unfortunately, during the third \acp{VAULT} \Lyalpha\ flight on July 7, 2005
the shutter malfunctioned while the \acp{DOT} was co-pointed during
good seeing. 
The \acp{DOT} mosaic in Fig.~\ref{fig:halya} was taken the next
morning.
The recent CLASP-1 rocket flight
(\citeads{2016SPD....4710107K}) 
yielded faster-cadence but lower-resolution \Lyalpha\ images while the
\Halpha\ observing at the Dunn Solar Telescope suffered from clouds. 
Regretfully, the \Halpha--\Lyalpha\ comparison in Fig.~\ref{fig:halya}
remains the best there is.

\subsection{Conclusion of this section}
My conjecture to understand Fig.~\ref{fig:halya} is that \Lyalpha\
shows \acp{PHE}s as bright grains and the initial aftermaths of
near-horizontally launched ones as short fibrils, whereas \Halpha\
shows their subsequent cooling tracks as long contrail fibrils with
\acp{PSBE}-defined contrasts.
\rev{For such post-hot cooling features the Lyman lines show the
present whereas the opacities in the Balmer lines and the \HI\,ff
continuum are defined by the hotter past.}

\section{Discussion}\label{sec:discussion}

\subsection{Shock and post-shock visibilities}
Figure~\ref{fig:extalma} suggests that the denser HION shocks
become optically thick and therefore visible in \Halpha. 
The subsequent \acp{PSBE} lag suggests that cooling aftermath
clouds are also visible in \Halpha.

The actual existence of a HION-like shock-ridden domain in the
internetwork areas of the solar atmosphere was established
over two decades ago from \CaII\,\HtwoV\ grains (\eg\
\citeads{1991SoPh..134...15R};
\citeads{1994chdy.conf...47C}; 
\citeads{1997ApJ...481..500C}); 
I then called it ``clapotisphere''in a review
(\citeads{1995ESASP.376a.151R}). 
It is obvious in all internetwork areas in high-resolution \CaIIH\
filtergram movies, \eg\ those on my \wldotmovies{DOT movie page}
and on the \href{http://hinode.nao.ac.jp/QLmovies} {Hinode
quick-look movie pages} and in ultraviolet continuum movies (\eg\
\citeads{2001A&A...379.1052K}), 
but it is very hard to detect in \Halpha\
(\citeads{2008SoPh..251..533R}) 
in contrast to the 7000\,K extinction expectation from
Fig.~\ref{fig:extalma}.
I attribute this surprising paucity of both shocks and
cooling aftermaths to obscuration by overlying \Halpha\ contrail
fibrils in which the gas has experienced higher
temperatures than in shocks and retains \acp{PSBE}
non-transparency.

The similarities between \Halpha\ and the \HI\,ff continua in
Fig.~\ref{fig:extalma} suggest comparable visibility and obscuration
of clapotispheric shocks and their cooling aftermaths in \acp{ALMA}
diagnostics.

\subsection{Detection of  contrail  precursors}
My suggestion that most if not all long \Halpha\ fibrils are contrails
requires ubiquitous precursor \acp{PHE}s that have not yet been
identified, but may be similar to the small fast heating events of
\citetads{2011Sci...331...55D}, 
\citetads{2011ApJ...738...18T}, 
\citetads{2015ApJ...814..123S}, 
\citetads{2016A&A...589A...3S}. 

The hot precursor of the contrail fibril of \PubII\ was visible as an
extending bright streak in 1400\,\AA\ slitjaw images from \acp{IRIS},
in yet hotter diagnostics from \acp{SDO}/\acp{AIA}, and in the far
blue wing of \Halpha\ as a dark streak due to large blueshift
and thermal core broadening.
It may have gotten its joint visibility in these diverse diagnostics
by being relative opaque, large, and slow. 
Smaller events may lose visibility from smaller optical
thickness or insufficient angular resolution.

The question so arises which diagnostics suit best to spot smaller,
less opaque, possibly faster heating events than \acp{RBE}s,
\acp{RRE}s, and the contrail producer of \PubII.  
The \acp{SB} curves in Fig.~\ref{fig:extalma} suggest much larger
extinction coefficient for hot \acp{PSBE} precursors in
\Lyalpha\ and the mm wavelengths than for \Halpha, also larger
than in other chromospheric lines 
(\rev{\linkadspage{2016A&A...590A.124R}{9}{Fig.~5}} of \PubI).
This suggests that \Lyalpha\ and mm wavelengths are the best to find
them, with largest opacity and emissivity in \Lyalpha\ where
hydrogen remains partially neutral and largest opacity and emissivity
at mm wavelengths at full hydrogen ionization.
In the absence of a fast-cadence high-resolution \Lyalpha\ space
mission \acp{ALMA} is the most promising facility if it
reaches sufficient angular resolution.

Intrinsic solar-atmosphere scattering blurs such
precursors to larger apparent extent in \Lyalpha\ images, but
not at mm wavelengths so that for
\acp{ALMA} the required angular resolution to detect \acp{PHE}s
as small bright blobs or jets is higher than in \Lyalpha.
\acp{ALMA} will show them in the detail permitted by the array
resolution.  
At only partial hydrogen ionization \Lyalpha\ scattering
from the precursor \acp{PHE} into cooler surrounding gas will produce
dark opaque aureoles around such bright kernels in \acp{ALMA} images.

\subsection{Fibrils with ALMA}  
Hotter precursors than HION shocks will leave larger
\acp{PSBE} opacities since higher temperature
implies larger $\varepsilon$ in \Lyalpha.
The precursor in the contrail of \PubII\ must have reached
temperatures above 10\,000\,K since it was visible in UV
\acp{IRIS} and EUV \acp{AIA} images.
Near this temperature the \acp{SB} \Halpha\ curves reach their peak
while the \HI\,ff curves reach yet higher saturation levels.
If the postulated precursors become this hot or even hotter, then the
\acp{non-E}-retarded contrail opacity in the post-event
cooling phase will be much larger for the mm continua than for
\Halpha. 
I therefore expect mm contrail fibrils to be yet more opaque than the
\Halpha\ ones, constituting a yet denser canopy.

\subsection{Fibrils  with Bifrost}
The most elaborate studies of \Halpha\ fibril formation in
numerical simulations of the solar atmosphere are those by Leenaarts
\etal\
(\citeyearads{2012ApJ...749..136L}, 
\citeyearads{2015ApJ...802..136L}), 
using a snapshot of a 3D \acp{non-E} \acp{MHD} simulation with the
Bifrost code (\citeads{2011A&A...531A.154G}) 
that was later made public by
\citetads{2016A&A...585A...4C}. 
The same snapshot was used in predictions for \acp{ALMA} by
\citetads{2015A&A...575A..15L}.

This snapshot shows some \Halpha\ fibrils on the condition that the
spectral synthesis is done with 3D radiative transfer so that the
larger-contrast granulation signature imposed by thermal photon
creation in the deep photosphere is erased by
\revdel scattering within the overlying chromosphere
(\rev{\linkadspage{2012ApJ...749..136L}{7}{Fig.~7} of}
\citeads{2012ApJ...749..136L}). 
This implies that the Bifrost fibrils are less opaque than actual
solar fibrils which do not let even very bright Ellerman bombs shine
through at line center. 
Also, the \rev{synthetic} \Halpha\ image shows far fewer \Halpha\
fibrils than areas with a similar amount of magnetism would show in
actual observations. 
\rev{They mostly connect opposite-polarity network patches; none jut
out far over adjacent internetwork.} 
\revdel

I suspect that these differences are due to various deficiencies in
the Bifrost studies.
First, although the \acp{MHD} simulation accounted for non-equilibrium
hydrogen ionization, the \rev{\Halpha\ synthesis} did not by applying
statistical-equilibrium \acp{NLTE} on the single snapshot so that the
\Halpha\ opacities had no post-hot boosting.
\revdel
The relatively few Bifrost fibrils gained \Halpha\ and \acp{ALMA}
contrasts only from density differences, not from memorial opacities.
\rev{There was also no opacity spreading from \Lyalpha\ scattering
since the net radiative rates of the Lyman transitions were assumed
zero for tractability as in the HION atmosphere of
\citetads{2007A&A...473..625L}}. 

Second, the non-equilibrium increase of the electron density that
resulted from retarded hydrogen recombination in the simulation and
was retained in the snapshot analysis tends to force hydrogen back to
the \HI\ ground level, just as in \acp{SB} equilibrium, giving 
smaller \Halpha\ extinction than without such \acp{non-E}
electron density increase. 

Finally, while Bifrost simulations do an outstanding job in emulating
solar granulation, acoustic waves and shocks, network-like field
concentrations, dynamic fibrils, and more of the rich zoo of
solar-atmosphere phenomena, they have not yet produced spicules-II or
equivalent \acp{RBE}s and \acp{RRE}s.
It is not for me to elucidate why Bifrost fails to make these,
but I do speculate that Bifrost also does not make the smaller and
hotter \acp{PHE}s that I deem required to explain ubiquitous
\Halpha\ fibrils.

\rev{About} the same few \rev{fibrils} appeared in the
\rev{longer-wavelength synthetic} \acp{ALMA}-prediction images in
\rev{\linkadspage{2015A&A...575A..15L}{4}{Fig.~4} of}
\citetads{2015A&A...575A..15L}. 
\rev{These resulted from evaluation of the integral 
\acp{LTE} transfer equation in terms of brightness temperature using
\HI\,ff opacities given by the \acp{non-E} electron and proton
densities in the simulation. 
However, long fibrils that extend away from the network areas
are also missing in these synthetic images.}

Since the actual Sun is largely covered by \Halpha\ fibrils and
Fig.~\ref{fig:extalma} predicts that these will be as opaque at mm
wavelengths in the case of instantaneous opacities and much
more opaque in the case of \acp{PSBE} memories, I wouldn't trust any
fibril-lacking simulation to suggest what \acp{ALMA} will observe
in fibrilar areas. 

\subsection{Fibril widths}
Resonance scattering of the intense \Lyalpha\ radiation from
recombining gas along \acp{PHE} precursor tracks boosts \Halpha\ and
\HI\ free-free extinction over a few hundred km in cooler gas around
the tracks (\rev{\linkadspage{2016A&A...590A.124R}{6}{Fig.~3}} of
\PubI, \linkssxp{26}{aureole boosting}).
Contrail fibrils therefore show widths of this extent in \Halpha\ even
when the actual precursors are much smaller and harder to detect --
just as aircraft jet engines are smaller on the sky and harder to
observe than the contrails they produce. 
\acp{ALMA} images will show them opaque over similar or yet larger
width, but render the actual cross-section temperature profile as
brightness.

\subsection{Fibril temperatures}
I base a rough estimate of contrail fibril temperatures on a
comparison of the \CaIR\ and \Halpha\ \acp{SB} extinction curves in
Fig.~\ref{fig:extalma} with the observational \CaIR--\Halpha\
comparison for a quiet-Sun scene in
\citetads{2009A&A...503..577C}. 
It showed that fibrils appear similarly in both lines near network,
but in their jutting out across internetwork they quickly become
transparent in \CaIR\ whereas they extend further out in
\Halpha. 
As a result, the brightness-brightness scatter plot in the first
column of \rev{\linkadspage{2009A&A...503..577C}{7}{Fig.~6}} of that
article
shows only correlation for the brightest network samplings.
However, the same figure shows high correlations between \Halpha\ core
width, \CaIR\ core width, and \CaIR\ Dopplershift-following
line-minimum intensity were these quantities are not small.
\citetads{2009A&A...503..577C} 
attributed these good correspondences to joint temperature
sampling.
For the \Halpha\ core width this was expected because
its thermal broadening is relatively large due to the small atomic
mass of hydrogen; for \CaIR\ core width and intensity it was not
surprising since both are temperature-sensitive.
The maps of these quantities in Figs.~3 and 4 in that article then
suggest lower fibril temperatures further away from network. 

The \Halpha\ and \CaIR\ curves in \revdel
Fig.~\ref{fig:extalma} cross \rev{at 6000--7000\,K}, implying that
where chromospheric fibrils appear similar in these lines
their temperature is about this value.
This so becomes my prediction for temperatures to be measured
with \acp{ALMA} {in the initial parts of long fibrils}.
At these high temperatures \acp{non-E} retardation is small.
Adjacent-fibril contrast in line-core images comes from different
densities, different temperatures in the case of \CaIR, and especially
from different Dopplershifts with larger sensitivity for the
relatively narrow core of \CaIR.

Since further away from activity the actual fibrils become
cooler, the \CaIR\ and \Halpha\ opacities along them get smaller in
the case of instantaneous \acp{SB} population following the leftward
declines in Fig.~\ref{fig:extalma}.
This decline is much steeper for \Halpha\ due to its large Boltzmann
sensitivity, so that in quiescent conditions \Halpha\ fibrils should
become transparent and invisible well before \CaIR\ fibrils and
therefore should appear shorter -- instead of being longer as
observed. 

My remedy for this incongruity is again to
postulate that hotter events have passed previously, making the
\Halpha\ extinction coefficient initially much larger (up the
\Halpha\ slopes towards the peaks in Fig.~\ref{fig:extalma}) and that
the post-event cooling gas retains such high \Halpha\ extinction
for minutes while the \CaIR\ extinction adjusts
near-instantaneously to the decreasing temperature. 

The observed transparency of outer fibrils in \CaIR\ then
suggests actual fibril temperatures around 5000~K or less away from
network, with long retardation and \acp{PSBE} opacities in
\Halpha\ and at mm wavelengths.

The brightness temperatures in \acp{ALMA} data should directly
correspond to these temperature predictions. 
In \Halpha\ the corresponding brightness-temperature range is much
smaller, covering only 4000-4200\,K in
\rev{\linkadspage{2009A&A...503..577C}{7}{Fig.~6}} of
\citetads{2009A&A...503..577C}, 
because it is set by scattering. 
Nevertheless, contrail fibrils will have good dark--dark
correspondence in byte-scaled \acp{ALMA}--\Halpha\ brightness
comparisons because low fibril temperature translates directly into
low mm brightness but also indirectly into low \Halpha\ brightness via
large \acp{PSBE} opacity.

The arguments above for the \CaIR--\Halpha\ incongruity were
earlier used in
\citetads{2008SoPh..251..533R} 
to explain the lack of fibril-canopy signature at the center of
\CaIIH\ in the venerable spectrogram sequence of
\citetads{1993ApJ...414..345L}. 
Although these spectrograms became famous through the Radyn
internetwork oscillation emulation by
\citetads{1997ApJ...481..500C}, 
the original analysis concentrated on long-period network oscillations
at \CaIIH\ center that \rev{in hindsight} may have been a signature
of repetitive \rev{PHE} launching
\rev{seen as accelerating Dopplergram branches jutting out
from network in 
Fig.~2 of
\citetads{2008SoPh..251..533R},} 
possibly producing the thin \CaIIK\ canopy fibrils reported by
\citetads{2009A&A...502..647P} 
\rev{that seem to correspond to thin straws near the limb
(\linkadspage{2006ASPC..354..276R}{2}{Fig.~1} of
\citeads{2006ASPC..354..276R}) 
and off-limb spicules-II
(\citeads{2007PASJ...59S.655D}). 
}

Finally, it is impossible to quantify the temperature of the
postulated \acp{PHE} precursors because they have not yet been
identified. 
However, the visibility of the contrail precursor of \PubII\ in UV
\acp{IRIS} and EUV \acp{AIA} images suggests that such events may
become significantly hotter than 10\,000\,K. 

\section{Conclusion}  \label{sec:conclusion}
The demonstration in
\rev{\linkadspage{2016Msngr.163...15W}{4}{Fig.~4}} of
\citetads{2016Msngr.163...15W} 
that \acp{ALMA} ``can serve as a linear thermometer for the
chromospheric plasma'' is obviously correct, but I suggest that scenes
as the simulation-predicted clapotisphere depicted in that figure will
hardly be detected by \acp{ALMA}, just as they are hard to find in
\Halpha.

In summary, Figure~\ref{fig:extalma} shows that if \Halpha\
fibrils are instantaneous 7000\,K features (as is the ALC7
chromosphere) then their opacities \revdel will be similar to those
in \Halpha\ \rev{at 0.35\,mm}, \revdel larger at 1.3\,mm, and
\revdel much larger at 3.0\,mm.
If instead most long \Halpha\ fibrils represent
retarded-opacity features after heating events as the one in
\PubII, \ie\ are contrails as I suggest from the striking
scene difference in Fig.~\ref{fig:halya}, the paucity of \Halpha\
shock scenes and the fibril incongruity with \CaIR, then their
opacities are very much larger at all \acp{ALMA} wavelengths. 

\revdel My prediction is that in \acp{ALMA} images most of the solar
surface will be covered by opaque \Halpha-like fibrils.
I think it naive to ignore the observed \Halpha\ truth when basing
predictions for \acp{ALMA} on solar-atmosphere simulations that lack
the ubiquitous fibrilar \Halpha\ \rev{canopies}.

However, on a more positive note, the fibrilar chromosphere and
especially the proposed precursor heating events represent a much more
interesting and promising research topic than internetwork shocks
which are well understood since
\citetads{1994chdy.conf...47C} 
and do not play an important role in solar atmosphere heating
(\citeads{1995ApJ...440L..29C}). 

Because solar physics is a field of too scarce predictions I
summarize this study with a dozen specific ones:
\begin{enumerate} \vspace{0ex} \itemsep=1ex

\item although probably the Sun will be less active
when \acp{ALMA} starts solar observing, I predict that
most of the solar surface will be covered by long opaque fibrils in
\acp{ALMA} images;

\item more precisely, I predict that at the \acp{ALMA} wavelengths the
general appearance of the Sun will be similar to \Halpha\ images
with good dark--dark correspondence but with larger
fibril opaquenesses at mm wavelengths that increase with wavelength
\rev{and with less lateral fibril contrast due to insensitivity to
Dopplershifts};

\item yet more precisely, I predict that the actual fibril
temperatures fall into three categories: above 10\,000\,K in small
heating events propagating outward from activity,
around \rev{7000\,K} in the initial parts of resulting
fibrils, and cooling down to 5000\,K \rev{and} less along
subsequent long contrails emanating far across internetwork;

\item while \acp{ALMA} can easily quantify the second and
third temperature categories, the first consists
of difficult, hard-to-catch features. 
However, even while small and fast, such events have opacities
at mm wavelengths that are much larger than in \Halpha\ (at
high temperature even larger than in \Lyalpha), making them best
detectable in fast-cadence image sequences from \acp{ALMA}
if these reach sufficient angular resolution.
I optimistically predict that \acp{ALMA} will see them.  
Measuring their temperature, energy release, and
contribution to atmospheric heating is then an exciting
\acp{ALMA} quest: something new on the Sun -- more attractive than
detailing well-known non-heating acoustic shocks;

\item if \acp{ALMA} indeed detects such precursor heating events
then I predict that these initially possess darker opaque aureoles
from \Lyalpha\ scattering and that such sunny-side-up morphology
vanishes from complete hydrogen ionization within the precursor at
\acp{ALMA}-measured temperatures below 15\,000\,K, closer to
the \acp{SB} than the \acp{CE} limit;

\item I predict that even when the hot precursors are
very small they produce contrail widths of order 0.5\,arcsec through
\Lyalpha\ scattering;

\item I predict that the precursor events are better field mappers
from line tying while hydrogen gets ionized than subsequent
less tied and wider contrail fibrils;

\item I predict that \acp{ALMA} will only sample internetwork shocks
in rare, utterly quiet areas free of fibrilar obscuration and
there will also detect subsequent cooling clouds
with temperatures dropping below 4000\,K conform the
COmosphere of \citetads{1996ApJ...460.1042A};

\item I predict that \acp{ALMA} will not observe any Ellerman bomb. 
They will be as obscured by fibril canopies as at the center of
\Halpha;

\item more positively, I predict that \acp{ALMA} will observe more
forceful reconnection events that break through or occur above the
fibril canopy and contribute coronal heating, in particular the
flaring active-region fibrils described in
\citetads{2015ApJ...812...11V} 
and \PubI. 
So far they are best seen in ultraviolet continua but they can also
appear as IRIS bombs in ultraviolet lines
(\citeads{2014Sci...346C.315P}; 
\citeads{2015ApJ...812...11V}). 
It is of large interest to track their temperature;

\item I predict that spicules-II will be much more opaque
in off-limb imaging with \acp{ALMA} than in \Halpha\ and \CaIIH;

\item similarly, I predict that wherever coronal rain is opaque in 
\Halpha\ it will be much more opaque at mm wavelengths.

\end{enumerate}

\noindent
Hopefully these predictions will soon be verified with actual \acp{ALMA}
observations.
I look forward to be proven right or wrong.

I end this study with a speculation.  
Quiescent filaments, in particular threads in their barbs and
legs, may likewise obtain their extraordinary \Halpha\ visibility
from post-hot opacity produced by 
frequent small heating events.
They will then have yet larger opacities and show up thicker
in \acp{ALMA} images, at lower
temperatures than suggested by statistical-equilibrium \Halpha\ modeling.

\begin{acknowledgements}
I am much indebted to S.~Toriumi of the National Astronomical
Observatory of Japan for inviting me for an extended stay and being an
excellent and efficient host. 
This analysis resulted from illuminating and inspiring discussions
there with R.~Ishikawa, M.~Kubo, M.~Shimojo and T.J.~Okamoto.
I thank the referee for suggesting many presentation improvements.
\wlchianti{CHIANTI} is a project of George Mason University, the
University of Michigan and the University of Cambridge.
I made much use of the \wlsolarsoft{SolarSoft} and \acp{ADS}
libraries.
My LaTeX macro to make in-text citations link to \acp{ADS} was
improved by \acp{EDP} Sciences; 
\rev{the trick to link to specific pages came from E.~Henneken
at \acp{ADS}.}
The \acp{EDP} production of A\&A also conserves my definition popups for
acronyms. 
\end{acknowledgements}

\bibliographystyle{aa-note}
\bibliography{AA29238}  

\end{document}